\documentclass[aip,jcp,reprint,floatfix,amsfonts,amssymb,amsmath]{revtex4-1} 

\usepackage{amsmath}
\usepackage{xcolor}
\colorlet{mylinkcolor}{blue!66!black!80}
\usepackage[colorlinks=true,linkcolor=mylinkcolor,citecolor=mylinkcolor,filecolor=cyan,
urlcolor=mylinkcolor,breaklinks=true]{hyperref}
\usepackage{mathtools}
\usepackage[capitalize]{cleveref}
\usepackage{braket}
\usepackage{bbold}
\usepackage[version=4]{mhchem}
\newcommand{\e}{{\rm e}}

\begin{document}

\setcounter{page}{1} %first page number

\title{Stochastic thermodynamics of chemical reactions coupled to finite reservoirs: A case study for the Brusselator}
\author{Jonas H. Fritz}
\author{Basile Nguyen}
\author{Udo Seifert}
\email[For correspondence:]{useifert@theo2.physik.uni-stuttgart.de}
\affiliation{II. Institut für Theoretische Physik, Universität Stuttgart, 70550 Stuttgart, Germany}
\date{\today}%

\begin{abstract}
{
Biomolecular processes are typically modeled using chemical reaction networks coupled to infinitely 
large chemical reservoirs. A difference in chemical potential between these reservoirs can drive the 
system into a non-equilibrium steady state (NESS). In reality, these processes take place in finite systems containing a 
finite number of molecules. In such systems, a NESS can be reached with the help of an externally 
driven pump for which we introduce a simple model. Crucial parameters are the pumping rate and the 
finite size of the chemical reservoir. We apply this model to a simple biochemical oscillator, the 
Brusselator, and quantify the performance using the number of coherent oscillations. As a surprising 
result, we find that higher precision can be achieved with finite-size reservoirs even though 
the corresponding current fluctuations are larger than in the ideal infinite case.
}
{}{}
\end{abstract}

\maketitle 

\section{Introduction} \label{sec:introduction}

Biological systems require a constant supply of energy to perform their tasks. There are numerous 
examples from molecular motors \cite{howa01,schl03,phil09,albe15} such as kinesin or myosin to 
biological switches and oscillators such as the circadian clock \cite{gold96,naka05}, the MinDE 
system \cite{fisc10,hala12,xion15,wu16,denk18} or the interlinked GTPase cascade 
\cite{mizu12,suda13,beme15,ehrm19}.  All of these systems reach a non-equilibrium steady 
state (NESS) by extracting energy from nucleotide phosphates such as adenosine or guanosine 
triphosphate  (ATP or GTP) \cite{west87,kame13}. Chemical energy is released through a hydrolysis 
reaction which breaks one of the phosphate bonds (dephosphorylation) and converts a nucleotide 
triphosphate (ATP) into a nucleotide diphosphate (ADP, GDP) and an inorganic phosphate ($\text{P}_\text{i}$). The thermodynamic models which describe these systems generally rely on infinitely large reservoirs that supply these species at a fixed concentration, which are called chemostats. In reality, these processes take place in finite systems with a finite number of molecules.  As a consequence, real biological oscillators do not have infinitely big chemostats at their disposal.

Instead, living systems rely on cellular respiration, a metabolic process that recycles ADP into ATP \cite{rich03,stry12} in order to remain in a NESS. It starts with glycolysis which converts glycose into pyruvates, then a series of oxidative phosphorylation results in the pumping of protons across the mitochondrial membrane creating an electrochemical gradient. The final recycling step is performed by a rotary molecular motor, namely the ATP synthase or $\text{F}_0\text{F}_1\text{-ATPase}$ in bacteria \cite{boye97,jung15}. The main purpose of this molecular motor is to maintain a NESS for different processes in the cell. The $\text{F}_0$-part is embedded in the membrane and couples to the proton gradient to rotate a central shaft. The $\text{F}_1$-motor uses ATP hydrolysis to rotate in the opposite direction. By coupling the two parts with a strong enough proton gradient, the $\text{F}_1$-motor rotates in reverse and thereby synthesizes ATP from ADP and $\text{P}_\text{i}$. Experimental techniques have enabled the observations of individual trajectories at the single molecule level \cite{toya10,toya11,toya15}. From such trajectories, efficiencies for the motor could be computed \cite{gasp07,gerr10,zimm12}. 

In this paper, we replace the chemostats with finite reservoirs and add a simple reaction scheme that fulfills a similar role to the $\text{F}_0\text{F}_1\text{-ATPase}$. Reactions in which a molecule leaves or enters the finite reservoir will change its concentration depending on the bath size. We quantify this effect with a parameter $\Lambda$ which describes how large the bath is compared to the rest of the system. The change in concentration is inversely proportional to the bath size, so in the limit $\Lambda \rightarrow \infty$ the change vanishes and we recover the ideal reservoir. We choose a simple unimolecular driven reaction with reaction speed $\gamma$ to mimick the role of the $\text{F}_0\text{F}_1\text{-ATPase}$ in cells. This reaction upholds the chemical free energy difference between the reservoirs. In a different context, a finite-size temperature bath has been considered in \cite{carc11,carc16} where it was modeled as a set of independent linear oscillators. 

We investigate how these modifications impact biochemical oscillators by considering the Brusselator model as a case study for how finite chemical reservoirs affect the performance of biological systems. 
Naively, one may expect that finite reservoirs would introduce additional noise into the system 
and lead to a decrease of its quality. We show that a simple biochemical oscillator with finite reservoirs 
can achieve higher precision than its counterpart with ideal reservoirs despite the increase in 
fluctuations. The relation between the precision of biochemical oscillations and the energy required 
to sustain them has received much attention recently for ideal reservoirs 
\cite{qian00,cao15,bara17,nguy18,fei18,herp18,mars19,zhan20, junc20, junc20a}.

This paper is organized as follows. In \cref{sec:model}, we introduce a Brusselator model and its 
modified version with a pumping mechanism. In \cref{sec:results}, we show that higher precision can 
be achieved with finite reservoirs despite showing larger fluctuations. We find that there is an optimal 
reservoir size and pumping speed which outperforms the ideal reservoir case.  We conclude in 
\cref{sec:conclusion}.

\section{Models} \label{sec:model}
\begin{figure}
     \includegraphics[scale=0.8]{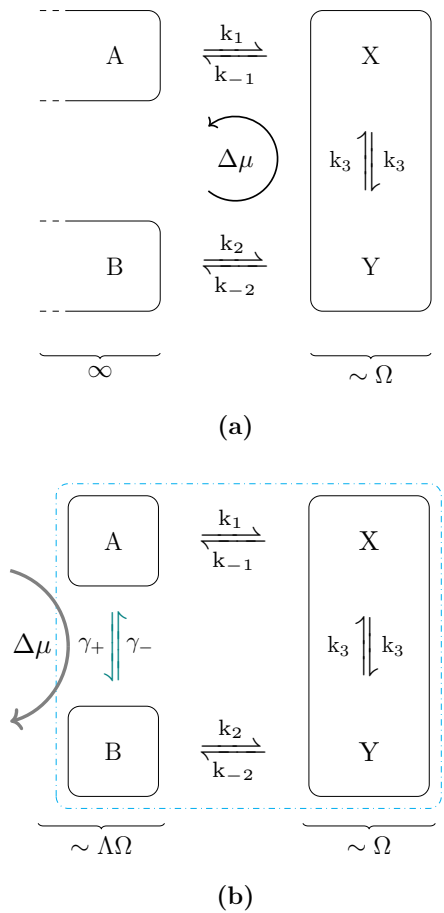}
     \caption{Comparison of the Brusselator (a) and our modified version with finite reservoirs (b). 
In (b) species $\mathrm{A}$ and $\mathrm{B}$  are part of the system. Their finite size scales 
with both with the system size $\Omega$ and $\Lambda$ which quantifies how large the size of the 
reservoirs is compared with the system size $\Omega$. Ideal reservoirs are recovered in the limit 
$\Lambda\rightarrow\infty$. To uphold the NESS an additional driven reaction scheme $\gamma_\pm$ highlighted in green is 
introduced. The dashdotted blue box separates the inside of the finite system from the outside from where the pump is fueled.}
\label{models}
 \end{figure}

 \subsection{The Brusselator model}
 The Brusselator is arguably the simplest set of chemical reactions that can exhibit oscillations as sketched in \cref{models}\,(a). It consists of two chemical species $\mathrm{X}$ and $\mathrm{Y}$ in a volume 
$\Omega$. $\mathrm{X}$ and $\mathrm{Y}$ molecules can be produced from chemostats containing 
$\mathrm{A}$ 
and $\mathrm{B}$ molecules, respectively. The set of chemical reactions is
\begin{equation}
\begin{aligned}
 \ce{A &<=>[k_1][k_{-1}] X}\, ,\\
 \ce{B &<=>[k_2][k_{-2}] Y}\, ,\\
 \ce{2X + Y &<=>[k_3][k_{3}] 3X}\, ,\\
\end{aligned}
\label{reactions}
 \end{equation}
 where the $k_1, k_{-1}, k_{2}, k_{-2}$ and $k_3$ are transition rates. This reaction scheme, first considered in \cite{qian02} is a modified version of the original Brusselator \cite{nico77,lefe88}. Through a chemical free energy difference $\varDelta\mu$ between the chemostats the system is driven out of equilibrium into a NESS. Thermodynamic consistency requires the rates $k_{i}$ and 
$\varDelta\mu\equiv\mu_{\mathrm{B}}-\mu_{\mathrm{A}}$ to be related via the 
local detailed balance condition, which in this case reads
\begin{equation}
    \mathrm{e}^{\varDelta\mu}=\frac{[\mathrm{B}] k_2 k_{-1} }{[\mathrm{A}] k_{-2} k_1} \, ,
\end{equation}
where we set $\beta=1$ throughout this paper\cite{seif12}. The thermodynamic force $\varDelta\mu$ must be above a 
certain critical threshold $\varDelta\mu_{\mathrm{c}}$ for biochemical oscillations to set in
\cite{cao15,nguy18}. 
% \[\]
 
\subsection{The Brusselator with finite reservoirs}
 
In the model with infinite reservoirs, the concentrations $[\mathrm{A}]$ and $[\mathrm{B}]$ remain constant, 
i.e., are chemostatted. For finite reservoirs this is no longer the case. The number of $\mathrm{A}$ and $\mathrm{B}$ 
molecules will now be part of the system and changes according to the set of chemical reactions. We 
introduce a parameter $\Lambda$, which characterizes the size of the reservoirs compared with the 
system size $\Omega$. The initial 
number of molecules in the bath is given by \eqref{reactions}
\begin{equation}
    \begin{aligned}
        n_{\mathrm{A}} = \left[\mathrm{A}\right]\Lambda\Omega\, ,\\
        n_{\mathrm{B}} = \left[\mathrm{B}\right]\Lambda\Omega\, ,\\
    \end{aligned}
\end{equation}
where $\left[\mathrm{A}\right]$ and $\left[\mathrm{B}\right]$ are the chemostatted concentrations from the original 
model. The total number of molecules 
$N_\text{tot}=n_{\mathrm{X}}+n_{\mathrm{Y}}+n_{\mathrm{A}}+n_{\mathrm{B}}$ becomes a conserved 
quantity.

In such a system, the NESS can be maintained by an externally driven pump. This driving has to be supplied by external ideal reservoirs, which are outside the dashdotted blue box in \cref{models}\,(b). In the case of the ATP synthase, the ideal reservoirs correspond to the the proton gradient. The NESS is reached by the pump sustaining a chemical gradient between $\mathrm{A}$ and $\mathrm{B}$. The simplest possible reaction scheme achieving this feature is 
\begin{equation}
 \ce{A <=>[$\gamma_+$][$\gamma_-$] B} ,
 \label{compensation}
\end{equation}
where $\gamma_\pm$ are transition rates, see \cref{models}\,(b). With this additional reaction, the 
local detailed balance condition reads
\begin{equation}
    \mathrm{e}^{\varDelta\mu}=\frac{\gamma_+ k_2 k_{-1}}{\gamma_- k_{-2} k_1} \, .
    \label{detailed balance}
\end{equation}
We assume that the rates $k_{\mathrm{i}}$ are fixed, thus  \cref{detailed balance} relates the 
ratio 
of $\gamma_+$ and $\gamma_-$ to a given $\varDelta\mu$. As a free parameter, we choose 
$\gamma\equiv\gamma_+$, which is the characteristic timescale of the pump. Our modified model is 
then described by three parameters: the size ratio $\Lambda$, the speed 
of the pump $\gamma$ and the chemical free energy difference $\varDelta\mu$.

\subsection{Chemical master equation and deterministic equations} 
The evolution of the probability to find the system in a state 
$\mathbf{n}\equiv\left(\mathrm{n_{\mathrm{X}}}, n_{\mathrm{Y}}, n_{\mathrm{A}}, 
n_{\mathrm{B}}\right)$ at time $t$ is described by the chemical master equation (CME) \cite{vank07}
\begin{widetext}
\begin{equation}
\begin{aligned}
&\partial_t P(\mathbf{n},t) =\Big\{k_1 \left[(n_{\mathrm{A}}+1) 
\epsilon_{\mathrm{X}}^-\epsilon_{\mathrm{A}}^+ - 
n_{\mathrm{A}}\right]  +  k_2 
\left[ (n_{\mathrm{B}}+1) \epsilon_{\mathrm{Y}}^-\epsilon_{\mathrm{B}}^+ - n_{\mathrm{B}}\right]  \\
& +  k_{-1}\left[(n_{\mathrm{X}}+1)\epsilon_{\mathrm{X}}^+\epsilon_{\mathrm{A}}^- - 
n_{\mathrm{X}}\right]  + 
 k_{-2}\left[(n_{\mathrm{Y}}+1)\epsilon_{\mathrm{Y}}^+\epsilon_{\mathrm{B}}^- - 
n_{\mathrm{Y}}\right] \\
& + \frac{k_3}{\Omega^2} 
\left[(n_{\mathrm{X}}-1)(n_{\mathrm{X}}-2)(n_{\mathrm{Y}}+1)\epsilon_{\mathrm{X}}^-\epsilon_{\mathrm
{Y}}^+ + 
(n_{\mathrm{X}}+1)n_{\mathrm{X}}(n_{\mathrm{X}}-1)\epsilon_{\mathrm{X}}^+\epsilon_{\mathrm{Y}}
^-\right. \\
& - n_{\mathrm{X}}(n_{\mathrm{X}}-1)n_{\mathrm{Y}}  - \left. 
n_{\mathrm{X}}(n_{\mathrm{X}}-1)(n_{\mathrm{X}}-2)\right] \\
& +\gamma^+\left[(n_{\mathrm{A}}+1)\epsilon_{\mathrm{A}}^+\epsilon_{\mathrm{B}}^- - 
n_{\mathrm{A}}\right] 
+\gamma^-\left[(n_{\mathrm{B}}+1)\epsilon_{\mathrm{B}}^+\epsilon_{\mathrm{A}}^- - 
n_{\mathrm{B}}\right] \Big\} P(\mathbf{n},t)\, ,
\end{aligned}
\label{eq:bruss_new_CME}
\end{equation}
\end{widetext}
where we define the step operators as
\begin{equation}
    \begin{aligned}
    \epsilon_{\mathrm{X}}^\pm P(\mathbf{n},t)&\equiv 
P(n_{\mathrm{X}}\pm1,n_{\mathrm{Y}},n_{\mathrm{A}},n_{\mathrm{B}},t)\, , \\
    \epsilon_{\mathrm{Y}}^\pm P(\mathbf{n},t)&\equiv 
P(n_{\mathrm{X}},n_{\mathrm{Y}}\pm1,n_{\mathrm{A}},n_{\mathrm{B}},t)\, , \\
    \epsilon_{\mathrm{A}}^\pm P(\mathbf{n},t)&\equiv 
P(n_{\mathrm{X}},n_{\mathrm{Y}},n_{\mathrm{A}}\pm1,n_{\mathrm{B}},t)\, , \\
    \epsilon_{\mathrm{B}}^\pm P(\mathbf{n},t)&\equiv 
P(n_{\mathrm{X}},n_{\mathrm{Y}},n_{\mathrm{A}},n_{\mathrm{B}}\pm1,t)\, . \\
    \end{aligned}
\end{equation}

In the deterministic limit 
($\Omega\rightarrow\infty$), we obtain from \cref{eq:bruss_new_CME} the rate equations for 
the concentrations,
\begin{equation}
    I \equiv \sum_{\mathbf{n}} n_I P(\mathbf{n},t)/\Omega\qquad I\in\{\mathrm{X, Y, A, B}\}.
\end{equation}
as
\begin{equation}
 \begin{aligned}
 \dot{X} &= k_1 A - k_{-1}X + k_3 (X^2Y-X^3)\, ,\\
 \dot{Y} &= k_2 B -k_{-2}Y + k_3(X^3-X^2Y)\, ,\\
 \dot{A} &= -k_1 A + k_{-1}X - \gamma_+ A + \gamma_- B\, ,\\
 \dot{B} &= -k_2 B + k_{-2}Y + \gamma_+ A -\gamma_- B\, .
 \end{aligned}
 \label{eq:deterministic_rate_equations}
\end{equation}

\subsection{Stochastic simulations}\label{sec:stochastic}
We have performed continous time Monte Carlo simulations of \cref{eq:bruss_new_CME} using 
Gillespie's algorithm \cite{gill77}. For all simulations we set the parameters to $\Omega = 1000$, 
$k_1=0.1/\Lambda$, $k_{-1}=0.1$, $k_2=0.1/\Lambda$, $k_{-2}=0.003$, $k_3= 0.001$ and 
$N_{\mathrm{tot}}=7.5\cdot\Omega\cdot\Lambda$. Here, $\Lambda$ is dimensionless parameter that reflects 
how large $n_\mathrm{A}+n_\mathrm{B}$ is compared to $n_\mathrm{X}+n_\mathrm{Y}$. 
We include $\Lambda$ in the total number of molecules. We also choose to scale the rates 
$k_1$ and $k_2$ with a factor $1/\Lambda$ in order to make the reaction 
propensities $k_1n_{\mathrm{A}}$ and $k_2n_{\mathrm{B}}$ independent of $\Lambda$. This ensures that we recover 
the ideal reservoirs dynamics in the limit $\Lambda\rightarrow\infty$. We choose $\varDelta\mu$, $\gamma$ and 
$\Lambda$ as control parameters. 

\begin{figure}
    \includegraphics[scale=0.4]{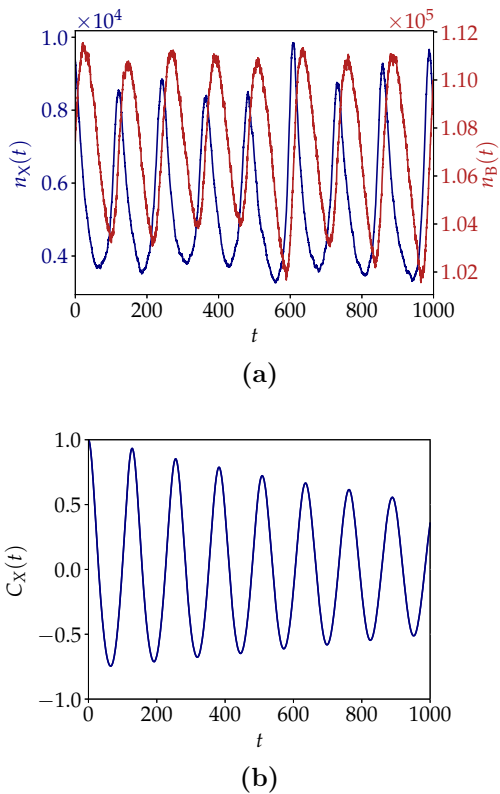}
    \caption{Stochastic trajectories of the Brusselator with finite reservoirs. (a) Trajectory of the number of X molecules $n_{\mathrm{X}}$ (blue) and B molecules 
$n_{\mathrm{B}}$ (red) where we set $\Lambda=20$, $\varDelta\mu =5.5$, $\gamma =1$. The remaining 
parameters are given in the main text. (b) Correlation function, fitted with \cref{eq:correlation_function}, 
leading to $\mathcal{N} \simeq 10$ coherent oscillations.}
     \label{correlation_decay}
\end{figure}

\subsection{Precision of oscillations and diffusion coefficient}\label{sec:precision_diffusion}

 In \cref{correlation_decay}\,(a), we plot an example of an oscillating trajectory for species $\mathrm{X}$ 
and $\mathrm{B}$. Such oscillations occur in systems with a finite number of molecules that display large 
fluctuations. To quantify the precision of oscillations, we compute the correlation function for 
species $\mathrm{X}$ which is defined 
as
\begin{equation}
\begin{aligned}
     C_{\mathrm{X}}(t) &\equiv 
\frac{\big\langle\left(n_{\mathrm{X}}(t)-\langle n_{\mathrm{X}}\rangle\right )\left( n_{\mathrm{X}}
(0)-\langle n_{\mathrm{X}}\rangle\right) \big\rangle\,}{\langle n_{\mathrm{X}}^2\rangle - \langle n_{\mathrm{X}}\rangle^2} \\
&\approx \exp\left(-t/\tau_c\right)\cos\left(2\pi t / T\right),
\label{eq:correlation_function}
\end{aligned}
\end{equation}
where the bracket denote an average over stochastic trajectories. We obtain the correlation time $\tau_c$
and the period length  $T$ by fitting the correlation function numerically with the function given in the second line of \cref{eq:correlation_function}. In \cref{correlation_decay}\,(b), we show a typical correlation function in 
the oscillating regime. The number of coherent oscillations is defined as the correlation time 
divided by 
the period length, i.e., 
\begin{equation}
 \mathcal{N}\equiv \frac{\tau_c}{T}.
 \label{eq:number_coherent}
\end{equation}
It measures how long different realizations of the oscillator stay coherent with each other. Thus $\mathcal{N}$ is a natural choice to quantify the precision of biochemical oscillations \cite{bara17,qian00}.

As a measure for fluctuations, we consider the current conjugated to the 
thermodynamic force $\varDelta\mu$, which is related to the entropy production. In the model with infinite reservoirs, 
this current is given by the rate of consumption of $\mathrm{B}$. In our modified 
model, the thermodynamic flux is related to the pumping scheme \eqref{compensation}. We can analyse 
its fluctuations by considering the stochastic time-integrated current $Z$. In a stochastic 
trajectory,  this random variable increases by one if a $\mathrm{A}$ is converted to an $\mathrm{B}$ 
molecule, which happens if the transition with rate $\gamma_+$ takes place. Likewise $Z$ decreases 
by one if an $\mathrm{B}$ is converted to a $\mathrm{A}$ molecule, which happens if the transition 
with rate $\gamma_-$ takes place, i.e.,
\begin{equation}
\begin{aligned}
\ce{A &->[\gamma_+]B}\qquad Z\rightarrow Z+1 \, , \\
\ce{B &->[\gamma_-]A}\qquad Z\rightarrow Z-1 \, .
\end{aligned}
\end{equation}
The average thermodynamic flux is then defined as
\begin{equation}
    J\equiv\lim_{\mathcal{T}\rightarrow\infty}\frac{\langle Z\rangle}{\mathcal{T}\Omega}\, .
\end{equation}
where $\mathcal{T}$ is the sampling time interval. Note that we choose our system parameters such that $J$ does not depend on the reservoir size $\Lambda$, this is due to the scaling factor $1/\Lambda$ in the rates $k_1$ and $k_2$. In the steady-state, the rate of entropy production is simply given by $\sigma\equiv J \varDelta\mu$. The fluctuations can be quantified by the diffusion coefficient, which is defined 
as
\begin{equation}
    D \equiv\lim_{\mathcal{T}\rightarrow\infty} \frac{\langle Z^2 \rangle-\langle Z\rangle ^2}{2 
\Omega\mathcal{T}}\, ,
    \label{eq:diffusion}
\end{equation}
At the onset of oscillations, this quantity diverges as a power-law with the system size 
\cite{nguy18}. Interestingly, $D$ has a universal lower bound that depends only on the thermodynamic 
force $\varDelta\mu$, which follows from the thermodynamic uncertainty relation 
\cite{bara15,ging16,seif19}. We choose $D$ as a quantifier for the fluctuations of the system.

\section{Results} \label{sec:results}
\subsection{Deterministic equations}
\begin{figure}
    \includegraphics[scale=0.35]{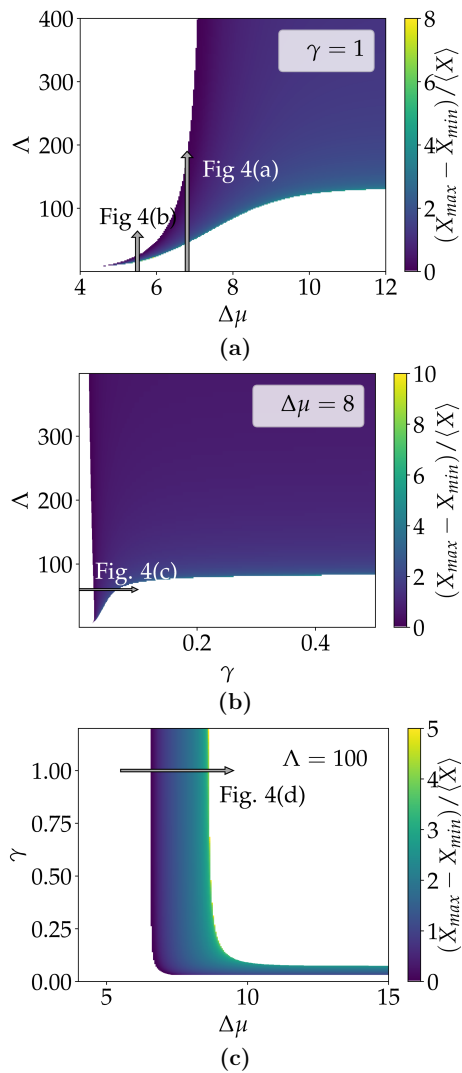}
    \caption{Three different cross sections of the $\varDelta\mu$-$\gamma$-$\Lambda$ phase-space.  
The amplitude of oscillations $\left(X_{\mathrm{max}}-X_{\mathrm{min}}\right)/\langle X\rangle$ is color coded. The white space indicates that no oscillations occur.}
    \label{det_phase_plots}
\end{figure}

We first consider the deterministic \cref{eq:deterministic_rate_equations} and study its 
non-equilibrium steady state solutions for which the left-hand side vanishes. These equations 
are solved numerically. We obtain the phase diagrams shown in \cref{det_phase_plots}, we plot the amplitude of oscillations defined as follows,
\begin{equation}
\textrm{Amplitude} = \frac{\mathrm{max}(X(t))-\mathrm{min}(X(t))}{\langle X\rangle}\equiv \frac{X_{\mathrm{max}}-X_{\mathrm{min}}}{\langle X\rangle}.
\label{eq:osc_amplitude}
\end{equation}
In all three cross sections there is an interplay between the parameters. For example, in \cref{det_phase_plots}\,(a), if the finite reservoir size ratio $\Lambda$ is too large for a fixed $\varDelta\mu$ the oscillations may vanish. A too large thermodynamic force $\varDelta\mu$ for a fixed $\Lambda$ can also make the oscillations vanish. We observe the same effect in \cref{det_phase_plots}\,(b) for $\gamma$, namely, a range of possible timescales for the pump that leads to oscillations. In \cref{det_phase_plots}\,(c) 
the interplay is most crucial: increasing both $\varDelta\mu$ and $\gamma$ too far simultaneously makes 
oscillations vanish as well.  Note that this effect, which is due to an additional fixed-point, is different from the case where $\varDelta\mu$ or $\gamma$ are too small. In the latter case, the system goes through a Hopf bifurcation, where the amplitude vanishes at the critical point. This difference can be seen qualitatively by the fact that the amplitude is different along the boundaries where oscillations vanish as will be explained in detail later.

\subsection{CME}
\begin{figure}
    \includegraphics[scale=0.39]{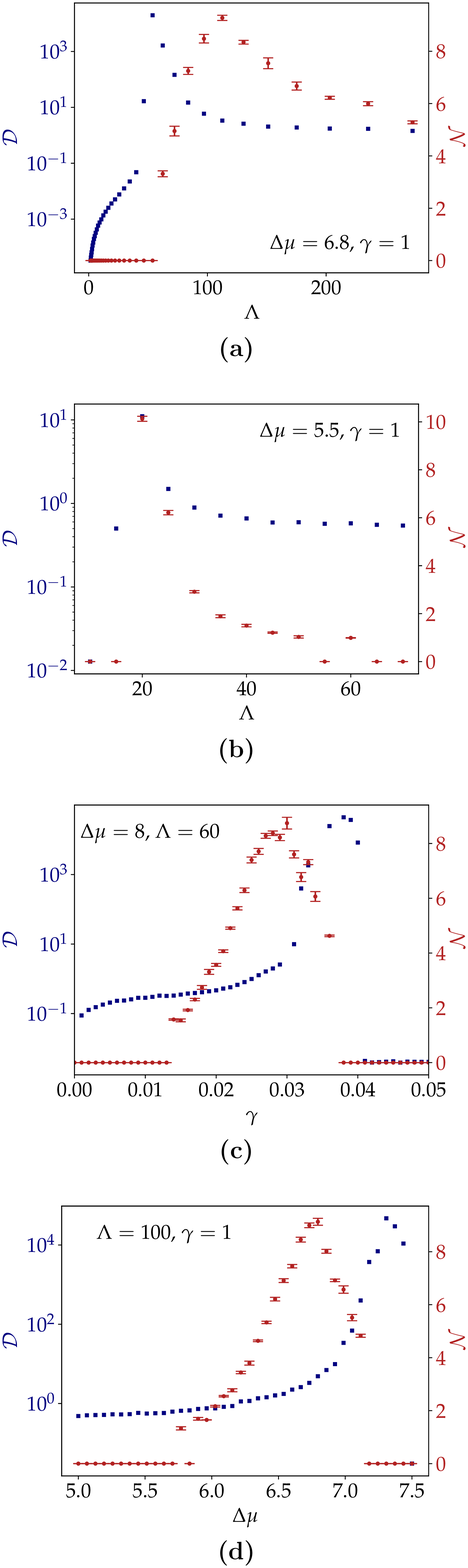}
\caption{Comparison between the diffusion coefficient (blue squares)  the number of coherent oscillations (red circles) as a function of the reservoir size $\Lambda$ for (a) and (b), pumping timescale $\gamma$ for (c) and thermodynamic force $\varDelta\mu$ for (d). The error bars result from fitting \cref{eq:correlation_function}.}
\label{CME_plots}
\end{figure}

We now consider oscillations in a system with a finite number of molecules that can display large 
fluctuations. We choose the diffusion coefficient $D$, \cref{eq:diffusion}, to quantify the 
fluctuations of the system and the number of coherent oscillations $\mathcal{N}$, 
\cref{eq:number_coherent}, to quantify the precision of oscillations. In \cref{CME_plots}, we plot 
the diffusion coefficient $D$ and the number of coherent oscillations $\mathcal{N}$  along the 
arrows shown in \cref{det_phase_plots}. At the phase transition, the diffusion coefficient has a 
local maximum \cite{nguy18}. Surprisingly, we observe that both 
$\mathcal{N}$ and $D$ increase simultaneously, in other words, higher precision can be achieved 
with stronger fluctuations. This is in contrast to the Brusselator model \cite{nguy18} and unicyclic models 
\cite{bara16,wier18}. The same feature, however, is shared by other models, such as the 
activator-inhibitor model\cite{nguy18}. In this sense, $\mathcal{N}$ and $D$ are not always 
strongly correlated in biochemical oscillators.

%In agreement with the deterministic results, we observe that $\Lambda$, $\gamma$ and $\varDelta\mu$  must be in a specific range to yield coherent oscillations. 
\begin{widetext}

\begin{figure}
    \includegraphics[scale=0.4]{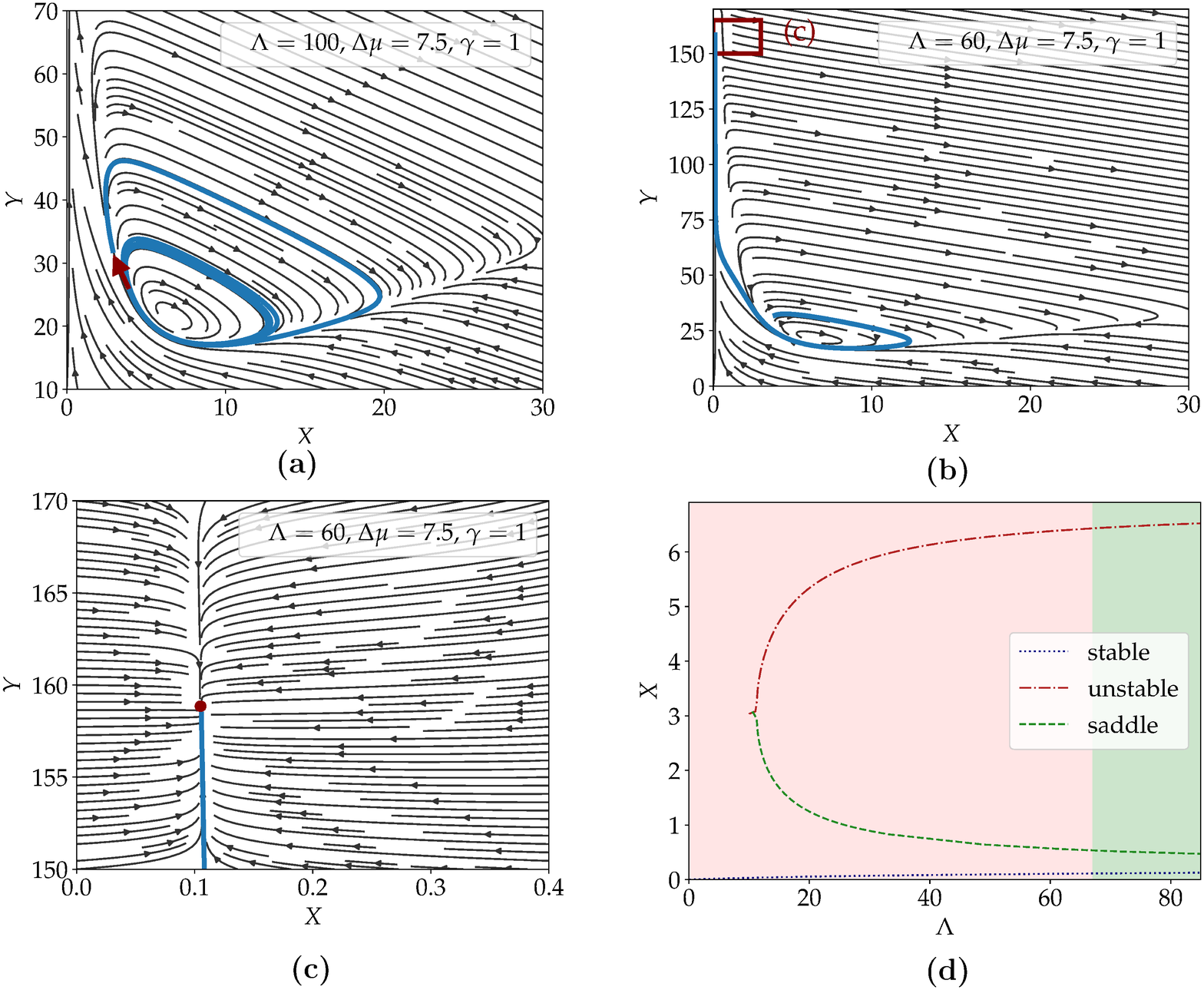}
\caption{Phase portrait in the $\mathrm{X}$-$\mathrm{Y}$ plane. (a) Emergence of additional larger 
cycles. We perturb the determininistic system along the red arrow. The system then goes through a 
larger cycle and converges back to the limit cycle (exemplary trajectory in blue). (b) Trajectory reaches a stable fixed-point located in the red square as $\Lambda$ decreases. (c) Enlargement of the red square from (b). It shows the locally stable fixed-point in red. (d) Bifurcation diagram. The fixed-points can be either stable (dotted blue), unstable (dashdotted red) or saddle points (dashed green). In the green region oscillations occur around the unstable point. At the red-green boundary a homoclinic bifurcation takes place.}
\label{stream_plots}
\end{figure}
\end{widetext}

What causes oscillations to vanish? First, $\varDelta\mu$ must be above a critical value $\varDelta\mu_c$ 
for oscillations to occur, which is also the case for the standard Brusselator. In our modified 
model, the threshold  $\varDelta\mu_c$ depends on the parameters $\Lambda$ and $\gamma$. Moreover, 
$\gamma$ and $\Lambda$ must be above certain thresholds for oscillations to set in; their specific 
critical values depend on the system parameters. In \cref{CME_plots}\,(a) and (b), we observe a 
decrease in the precision of oscillations for increasing $\Lambda$. As $\Lambda$ increases, so does 
$\varDelta\mu_c$, which causes $\mathcal{N}$ to decrease and, in \cref{CME_plots}\,(b), even to 
vanish. Most remarkably, we find that $\mathcal{N}(\Lambda)$ approaches 
$\mathcal{N}(\Lambda\rightarrow\infty)$, which corresponds to the infinite reservoir from above. 
This result implies that for values of $\varDelta\mu$ for which both models show oscillations, the oscillations driven by finite reservoirs can show higher precision than the ones with ideal reservoirs despite having larger fluctuations. In addition, the model with finite reservoirs can show oscillations when the one with ideal reservoirs does not. In this sense, the finite 
reservoirs outperform the infinite ones.

Second, for finite reservoirs, oscillations can vanish when $\varDelta\mu$ and $\gamma$ are too 
large, see \cref{det_phase_plots}\,(c). As shown in \cref{CME_plots}\,(c) and (d), $\mathcal{N}$ 
reaches a maximum and vanishes for large $\varDelta\mu$ and $\gamma$. For an explanation of this 
surprising effect, we plot the phase portrait of the deterministic system in \cref{stream_plots}. In 
the steady-state, the deterministic system will remain on a limit cycle as shown in 
\cref{stream_plots}\,(a). This is no longer the case for the stochastic system which can explore 
larger cycles due to fluctuations. We illustrate this by perturbing a deterministic 
trajectory located on the limit cycle as indicated by the red arrow. Where streamlines are dense, 
such a perturbation can lead to a large change in the trajectory. The system goes through a large 
cycle before converging back to the limit cycle. 

For the stochastic system, fluctuations are constantly perturbing the trajectory, which 
stochastically leads to large cycles. Their appearance results in a decrease in 
the number of coherent oscillations as these large cycles are no longer coherent with the 
oscillations in the limit cycle. As $\varDelta\mu$ and $\gamma$ are increased, transitions from the inner limit cycle onto a larger cycle are more likely 
to happen. The effect of varying $\Lambda$ is shown in \cref{stream_plots} (b)-(d). The bifurcation diagram \cref{stream_plots} (d) shows the $X$-value of fixed-point solutions of the deterministic system. In the green region for $\Lambda\gtrsim 67$, oscillations occur around the unstable fixed-point. As $\Lambda$ decreases below $\Lambda\simeq 67$, the system undergoes a homoclinic bifurcation \cite{strog01} through which the stable limit cycle disappears and trajectories end at a stable fixed-point as shown in \cref{stream_plots} (b) and (c).
It is interesting to note that in \cref{CME_plots} the largest coherence occurs closer to the homoclinic bifurcation than to the Hopf bifurcation. This is due to the fact that at the homoclinic bifurcation the oscillation amplitude is larger, making fluctuations less noticeable.
% Second, for finite reservoirs, oscillations can vanish when $\varDelta\mu$ and $\gamma$ are too large, see \cref{det_phase_plots} (c). As shown in \cref{CME_plots}\,(c) and (d), $\mathcal{N}$ reaches a maximum and vanishes for large $\varDelta\mu$ and $\gamma$. For an explanation if this suprising effect, we plot the phase portrait of the deterministic system in \cref{stream_plots}. With increasing $\varDelta\mu$ and $\gamma$, the increased likelihood of large cycles in the phase space of the system, as illustrated in \cref{stream_plots} (a). The determininistic system will stay on the inner limit cycle, whereas the stochastic system will enter one of the many larger cycles due to fluctuations, as highlighted by the red arrow. Fluctuations can lead to larger changes in concentrations along a trajectory in regions where the streamlines are dense. As soon as this happens, such a realization takes longer than a typical oscillation to return to the deterministic limit cycle and is thus no longer coherent with the other ones. As $\varDelta\mu$ and $\gamma$ are increased further, transitions from the inner limit cycle onto a larger cycle are much likely to happen. Finally, as $\Lambda$ decreases, an additional fixed point appears as depicted in \cref{stream_plots}\,(b). Trajectories that leave the limit cycle to this fixed point return on a much larger timescale.

For the parameter range plotted in \cref{CME_plots} as indicated in \cref{det_phase_plots}, 
the oscillations of $n_{\mathrm{X}}(t)$ and $n_{\mathrm{Y}}(t)$ are stabilized by oscillations in 
the number of $\mathrm{A}$ and $\mathrm{B}$ molecules in the reservoirs as shown in \cref{correlation_decay} (a). 
Oscillations can occur because the pumping mechanism \cref{compensation} does not attempt to keep 
the reservoir concentrations fixed. Rather, it restores the ratio of the bath concentrations to a 
fixed value given by $\varDelta\mu$, since $[\mathrm{A}]/[\mathrm{B}]$ is proportional to $\e^{-\varDelta\mu}$ 
according to the local detailed balance condition \cref{detailed balance}. At small bath scales 
$\Lambda$, from the systems perspective bath oscillations become noticeable and could qualitatively explain the 
improved precision of finite reservoirs in \cref{CME_plots}\,(a). 
% Similarly, we observe an optimal value for $\varDelta\mu$ and $\gamma$ in \cref{CME_plots}\,(c) and (d), where we believe that oscillations in the reservoirs match system oscillations.

\section{Conclusion} \label{sec:conclusion}

We have introduced a simple model for the role of finite reservoirs in chemical reaction networks. To uphold a NESS, we have introduced a pumping mechanism, which implicitly still assumes an ideal reservoir as a source of constant $\varDelta\mu$. The crucial point of our model, however, is that the number of A and B molecules required as a source for generating oscillations in X and Y are finite and fluctuating. This model thus better reflects the real conditions in biological systems than the usual assumption of infinite reservoirs. We have considered the simplest possible mechanism, a first-order chemical reaction converting species between reservoirs. This class of reservoirs is characterized by three parameters: the thermodynamic force $\varDelta\mu$, the bath scale $\Lambda$, which relates the reservoir size to the rest of the system and the timescale of the pumping mechanism $\gamma$. The 
ideal reservoirs are recovered in the limit $\Lambda\rightarrow\infty$ (independently of $\gamma$). 

As a case study, we have investigated a biochemical oscillator, the Brusselator, with this simple pumping mechanism. We quantify the precision of oscillations by measuring the number of coherent oscillations and the diffusion coefficient associated with the pump. We find that the occurence of oscillations critically depends on all of the parameters $\varDelta\mu$, $\gamma$ and $\Lambda$, in other words, oscillations can only occur if the thermodynamic force, the size of the chemostats and the pumping speed are within a certain range. Most surprisingly, the highest precision of oscillations occurs at finite parameters $\varDelta\mu$, $\gamma$ and $\Lambda$. This is in contrast to the Brusselator model \cite{nguy18} and other unicyclic models \cite{bara16,wier18} considered in the previous literature, where the precision of oscillations monotically increases with the control parameter $\varDelta\mu$. As a main result, we find that a system with finite reservoirs can outperform one with ideal reservoirs despite having larger fluctuations. 

Our framework could be extended by considering more sophisticated pumping mechanisms. It would be interesting to consider the ATP synthase, for which thermodynamically consistent models exist \cite{zimm12}. Specifically, it would be interesting to study the relation between the efficiency of the motor and the precision of oscillations. Another interesting case is the coupling of biochemical clocks which can be optimized to maximize the precision of oscillations \cite{zhan20}. Moreover, it has been shown recently that periodically driven oscillator can achieve better coherence than under NESS conditions \cite{ober19}. It would be interesting to investigate how fluctuations in the periodic protocol affect the precision of oscillations. Finally, we expect that considering finite reservoirs may show further surprises for biochemical systems beyond the enhanced precision of oscillations discovered here. 

% \section*{Acknowledgements}
% \appendix

% \clearpage
\section*{Data Availability Statement}
The data that support the findings of this study are available from the corresponding author upon 
reasonable request.

\end{document}